\documentclass[%
 reprint,
superscriptaddress,
amsmath,amssymb,
prl,
floatfix,
]{revtex4-2}

\usepackage{graphicx}
\usepackage{dcolumn}
\usepackage{bm}
\usepackage{color}
\usepackage[colorlinks=true,allcolors=blue,breaklinks=true]{hyperref}
\usepackage{physics}
\usepackage{xr}
\usepackage{float}
\makeatletter
\newcommand*{\addFileDependency}[1]{
  \typeout{(#1)}
  \@addtofilelist{#1}
  \IfFileExists{#1}{}{\typeout{No file #1.}}
}
\makeatother

\newcommand*{\myexternaldocument}[1]{%
    \externaldocument{#1}%
    \addFileDependency{#1.tex}%
    \addFileDependency{#1.aux}%
}
\myexternaldocument{supplement}

\begin{document}

\preprint{APS/123-QED}

\title{Mechanical Plasticity of Cell Membranes Enhances Epithelial Wound Closure}

\author{Andrew T. Ton}
\email{andrew.ton@yale.edu}
\affiliation{
 Department of Physics, Yale University, New Haven, Connecticut 06520, USA}
\affiliation{
 Integrated Graduate Program in Physical and Engineering Biology, Yale University, New Haven, Connecticut 06520, USA}
 
\author{Arthur K. MacKeith}
\affiliation{
 Department of Mechanical Engineering \& Materials Science, Yale University, New Haven, Connecticut 06520, USA}
\affiliation{
 Integrated Graduate Program in Physical and Engineering Biology, Yale University, New Haven, Connecticut 06520, USA}
 
\author{Mark D. Shattuck}
\affiliation{
 Benjamin Levich Institute and Physics Department, The City College of New York, New York, New York 10031, USA}

\author{Corey S. O'Hern}
\affiliation{
 Department of Mechanical Engineering \& Materials Science, Yale University, New Haven, Connecticut 06520, USA}
\affiliation{
 Department of Physics, Yale University, New Haven, Connecticut 06520, USA}
\affiliation{
 Department of Applied Physics, Yale University, New Haven, Connecticut 06520, USA}
\affiliation{
 Integrated Graduate Program in Physical and Engineering Biology, Yale University, New Haven, Connecticut 06520, USA}

\date{\today}

\begin{abstract}
During epithelial wound healing, cell morphology near the healed wound and the healing rate vary strongly among different developmental stages even for a single species like \textit{Drosophila}. We develop deformable particle (DP) model simulations to understand how variations in cell mechanics give rise to distinct wound closure phenotypes in the \textit{Drosophila} embryonic ectoderm and larval wing disc epithelium.  We find that plastic deformation of the cell membrane can generate large changes in cell shape consistent with wound closure in the embryonic ectoderm. Our results show that the embryonic ectoderm is best described by cell membranes with an elasto-plastic response, whereas the larval wing disc is best described by cell membranes with an exclusively elastic response. By varying the mechanical response of cell membranes in DP simulations, we recapitulate the wound closure behavior of both the embryonic ectoderm and the larval wing disc.
\end{abstract}

\maketitle
In response to wounding, epithelia carry out complex chemical and physical processes to restore tissue integrity. Epithelial wound healing has been studied in numerous species, including \textit{Drosophila}, zebrafish, and humans \cite{wood2002wound, abreu2012drosophila, tetley2019tissue, miskolci2019distinct, wong2012focal, guo2010factors, george2006basic}. Even within a single species, the healing process varies with developmental stage \cite{tetley2019tissue, redd2004wound}. In later stages, wound healing is slower, requires smaller changes in cell shape, and causes more scarring, which have been attributed to differences in chemical signaling, such as heightened inflammatory response \cite{redd2004wound, richardson2013adult, cordeiro2013role}. However, physical mechanisms, such as force transmission through cell junctions and collective cell motion, have also been shown to influence wound healing~\cite{li2013collective,brugues2014forces,ladoux2017mechanobiology}. An important driving force for wound closure across many developmental stages and species is the actomyosin purse-string that forms around the wound ~\cite{wood2002wound, redd2004wound, abreu2012drosophila, davidson2002embryonic, brugues2014forces}. Cell shape changes \cite{bi2015density,park2015unjamming,grosser2021cell} are another physical mechanism that can affect the dynamics of wound healing in epithelial tissues. An important open question is determining how these \textit{physical} mechanisms influence wound closure in different developmental stages. 

Previous computational models of wound closure have investigated contributions from substrate mechanical properties, active driving forces, and tissue tension \cite{bi2016motility,staddon2018cooperation,brugues2014forces, tetley2019tissue}. These models assume that cell membranes only respond elastically to deformation, ignoring viscoelastic and plastic response \cite{janshoff2021viscoelastic, shi2018cell, wyatt2016question}. However, recent experimental studies have shown that irreversible cell shape changes \cite{molnar2021plastic, staddon2019mechanosensitive, khalilgharibi2019stress, cavanaugh2020rhoa} are necessary for cell stress relaxation and tissue remodeling. Neglecting cell membrane plasticity can give rise to unrealistically large stresses when significant cell shape changes are required for wound closure.  It is therefore important to understand the role of viscoelastic and plastic response of cell membranes during wound closure. 

\begin{figure}[!t]
    \centering
    \includegraphics[width=\linewidth]{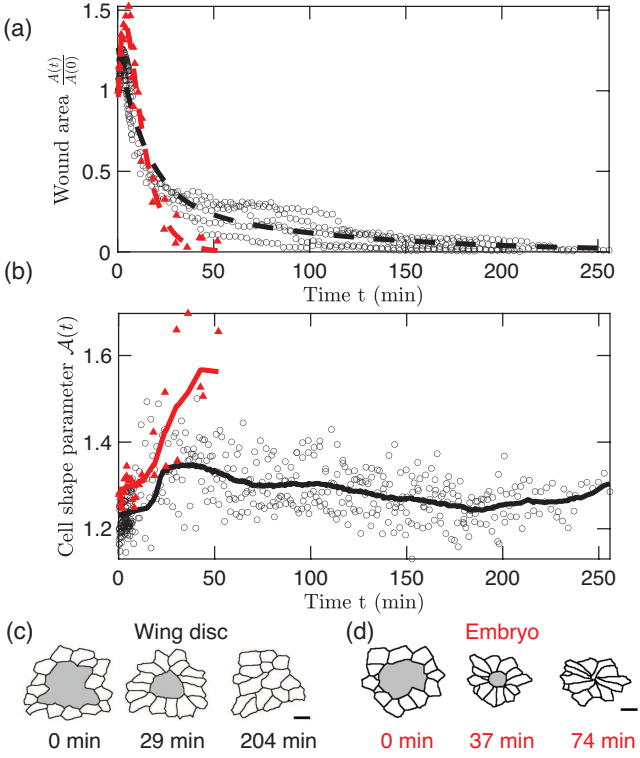}
\caption{(a) Normalized wound area $A(t)/A(0)$ plotted as a function of time $t$ during wound closure in wing discs (black circles) and embryos (filled red triangles), fit to a sum of two exponential functions (dashed lines). (b) Cell shape parameter $\mathcal{A}(t)$ averaged over cells adjacent to the wound boundary and plotted versus time during wound closure for the same data in (a). We include a moving average of the data in (b) with a $20$-minute window size (solid lines). The data for $A(t)/A(0)$ and $\mathcal{A}(t)$ are from $5$ wing disc wounds and $2$ embryo wound experiments conducted in Ref.~\cite{tetley2019tissue}. Example cell outlines reproduced from Ref.~\cite{tetley2019tissue} are shown during wound closure for a single (c) wing disc and (d) embryo, with the wound shaded in gray. The scale bars are 3 $\mu m$ in (c) and 5 $\mu m$ in (d). 
}
\label{fig:experimentalObservations}
\end{figure}

\textit{In vivo} studies of wound closure in late-stage \textit{Drosophila} embryonic ectoderm and late third-instar \textit{Drosophila} larval wing disc epithelium have found that cells near healed embryo wounds are elongated relative to those near healed wing disc wounds \cite{tetley2019tissue}. (See Fig.~\ref{fig:experimentalObservations}c,d.) Embryo wounds close at a rate of $\approx6.2~\mu $m$^2$/minute with cell shape changes of more than $30\%$ near the wound, whereas wing disc wounds close at a rate of $\approx0.7~\mu $m$^2$/minute with cell shape changes of less than $10\%$ near the wound.
(See Fig.~\ref{fig:experimentalObservations}a,b, plotted from experimental data in Ref. \cite{tetley2019tissue}.) The findings of Ref. \cite{tetley2019tissue} employ a vertex model approach \cite{alt2017vertex} to predict that greater cell intercalation rates lead to increased wound closure speed. This prediction leads to an open question of how, relative to wing disc wounds, embryo wounds have lower intercalation rates yet heal more quickly \cite{tetley2019tissue}. To address this question, we propose that embryo ectodermal cells can rapidly remodel their membranes to sustain greater cell shape changes, which leads to faster wound closure rates than in the wing disc epithelium. 

We carry out numerical simulations of the deformable particle (DP) model to explore the relationship between cell mechanical properties and wound closure phenotypes. We vary the degree of cell shape plasticity and determine the resulting effects on wound closure rate and cell shape deformation. We compare our simulation results to measurements of cell shape changes and wound closure rates from wounding experiments in embryonic and larval wing disc epithelia \cite{tetley2019tissue}. Our results suggest that cell shape plasticity is essential to achieve cell shape changes observed during embryo wound closure. Moreover, plasticity allows for faster wound closure rates in embryos compared to those in wing discs. 
\begin{figure}[!t]
    \centering
    \includegraphics[width=\linewidth]{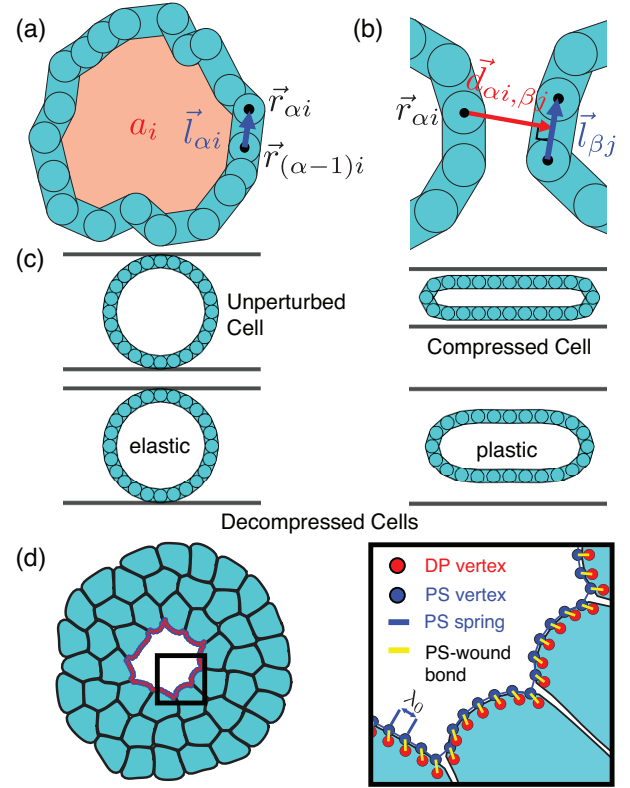}
\caption{(a) Schematic of a deformable particle (or cell) with area $a_i$ and segment lengths $l_{\alpha i} = |\vec{r}_{\alpha i}- \vec{r}_{(\alpha-1) i}|$ representing the cell membrane. (b) The distance vector ${\vec d}_{\alpha i,\beta j}$ between vertex $\alpha$ on cell $i$ and membrane segment ${\vec l}_{\beta j}$ on cell $j$ has no component along ${\vec l}_{\beta j}$. (c) A cell with $\mathcal{A}=1$ is uniaxially compressed. An elastic cell returns to its undeformed shape (left) after the strain is removed, whereas a plastic cell is permanently deformed (right). (d) A simulated wound is initialized as a cell monolayer, followed by removal of central cells such that the wound size is similar to those in Ref. \cite{tetley2019tissue}. Inset: Close-up of the purse string (PS), modeled as a collection of vertices (blue) along the edge of the wound (red). PS vertices are connected by springs (blue lines) with rest lengths $\lambda_0$, and each PS vertex is bonded to one DP vertex (yellow lines). 
}
\label{fig:modelSchematic}
\end{figure}

Since epithelial wound healing primarily involves in-plane motion, we consider a two-dimensional DP model for wound closure. The DP model has been studied recently in 2D and 3D, and DPs have been previously used to describe jamming and clogging of emulsion droplets and tissue morphogenesis \cite{boromand2018jamming, cheng2022hopper, wang2021structural, treado2021bridging, treado2022localized}. The strengths of the DP model include the ability to describe both confluent and non-confluent cell monolayers, both faceted and curved cell surfaces, and enable modeling both repulsive and cohesive intercellular forces. The shape energy \cite{boromand2018jamming} for each cell $i$ is 
\begin{equation}
    \label{eq:energy_DPM}
    U_{shape, i} = \frac{k_a}{2}\left(a_i - a_0\right)^2 + \frac{k_l}{2}\sum_{\alpha=1}^{N_v}\left(l_{\alpha i} - l_{0\alpha i}\right)^2 + U_{b,i}.
\end{equation}
Each cell is represented by a polygon with $N_v$ vertices (labeled by $\alpha$), membrane bond vectors $\vec{l}_{\alpha i} = \vec{r}_{\alpha i} - \vec{r}_{(\alpha-1) i}$, vertex positions ${\vec r}_{\alpha i}=(x_{\alpha i},y_{\alpha i})$, equilibrium area $a_0$, and equilibrium intervertex membrane length $l_{0\alpha i}$. The area stiffness spring constant $k_a$ and membrane length spring constant $k_l$ penalize deviations of the cell area $a_i$ from $a_0$ and membrane length $l_{\alpha i}$ from $l_{0\alpha i}$. (See Fig. \ref{fig:modelSchematic}a.) We quantify cell shape using the shape parameter $\mathcal{A}_i = p_i^2/4\pi a$, where $p_i$ is the perimeter of cell $i$ and ${\cal A}_i \ge 1$. The bending energy 
\begin{equation}
    \label{eq:bending}
    U_{b, i} = \frac{k_b}{2}\sum_{\alpha=1}^{N_v}\theta_{\alpha i}^2
\end{equation} 
determines the energy cost of membrane curvature for cell $i$, where $k_b$ is the membrane bending rigidity and $\theta_{\alpha i}$ is the angle between $\vec{l}_{\alpha i}$ and $\vec{l}_{(\alpha-1)i}$.

The cells interact through the pair potential $U_{int}$, which is a function of the distances between each vertex and nearby membrane segments on neighboring cells. $U_{int}$ includes soft-core repulsion and short-range attraction with a variable well depth. We calculate the distance 
\begin{align}
    \label{eq:point_line_distance}
    &d_{\alpha i,\beta j} = \\ 
    &\frac{|(x_{(\beta-1)j} - x_{\beta j})(y_{\beta j}-y_{\alpha i}) - (x_{\beta j} - x_{\alpha i})(y_{(\beta-1)j} - y_{\beta j})|}{|\vec{r}_{\beta j} - \vec{r}_{(\beta-1)j}|} \nonumber
\end{align}
between each vertex $\alpha$ on cell $i$ and membrane segment $\vec{l}_{\beta j}$ on cell $j$ as shown in Fig.~\ref{fig:modelSchematic}b. Having intercellular interactions that are only a function of $d_{\alpha i,\beta j}$ results in smooth sliding adhesion by eliminating components of the force on vertex $\alpha$ on cell $i$ from interactions with cell $j$ that are parallel to $\vec{l}_{\beta j}$. (See the definition of $U_{int}$ in Eq. S1, and the simulation parameters in Table S1 in Supplemental Material (SM) \cite{supp}.) The total energy $U_{DP}$ of a monolayer of $N$ cells is 
\begin{equation}
    U_{DP} = \sum_{i=1}^N U_{shape,i} + \sum_{i > j}^N \sum_{\alpha > \beta}^{N_v} U_{int}(d_{\alpha i,\beta j}).
\end{equation}

Animal cell membranes possess solid-like mechanical response on short and intermediate time scales, and are capable of stretching, bending, and transmitting forces \cite{dai1997cell, janmey2007cell}. To describe the viscoelasticity of the cell membrane, we model the membrane segments as springs that remodel their rest lengths in response to stress (Eq. \ref{eq:plastic_dynamical_eq}), similar in approach to models of irreversible deformation of the cytoskeleton and cell junctions \cite{munoz2013physiology, staddon2019mechanosensitive, cavanaugh2020rhoa}.  We use rest-length remodeling to describe the net result of membrane stress relaxation processes, such as actin cortex remodeling, membrane folding and unfolding via caveolae, and vesicle trafficking via endocytosis and exocytosis \cite{kelkar2020mechanics, figard2014membrane,cavanaugh2020rhoa}. We assume that the membrane segment rest length $l_{0\alpha i}$ obeys 
\begin{equation}
    \label{eq:plastic_dynamical_eq}
    \frac{dl_{0\alpha i}}{dt} = -\frac{k_l}{\eta}(l_{0\alpha i}-l_{\alpha i}),
\end{equation}
with damping coefficient $\eta$. The plastic relaxation timescale $\tau = \eta / k_l$ controls the membrane remodeling rate. In Fig. \ref{fig:modelSchematic}c, we show a compression test of duration $T$ on a single cell with an elastic ($\tau/T \gg 1$) and plastic membrane ($\tau/T \ll 1$). Cells with elastic membranes recover their undeformed shape, whereas cells with plastic membranes do not. By varying $\tau$, we describe cells with different degrees of elasto-plasticity. 

\begin{figure}[!th]
    \centering
    \includegraphics[width=\linewidth]{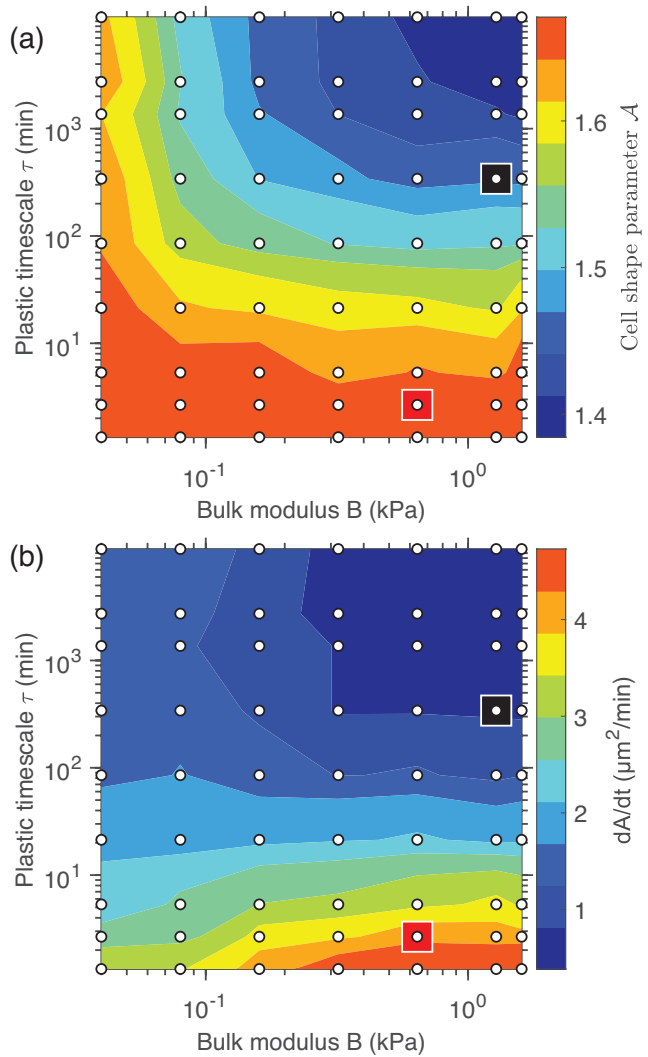}
\caption{(a) Cell shape parameter $\mathcal{A}$ in the healed system plotted versus the cell bulk modulus \textit{B} and plastic relaxation timescale $\tau$. (b) Closure rate $dA/dt$ plotted using the data in (a). Parameters indicated by the squares represent predictions for $B$ and $\tau$ of embryo (red) and wing disc (black) cells. $dA/dt$ is defined as $95\%$ of the maximum wound area divided by the time it takes for the wound to shrink to $5\%$ of its maximum size. Averaging at a given $\tau$ and $\beta$ is performed over $25$ simulations with different initial conditions. Simulations are carried out at $B$ and $\tau$ given by the grid points, and contours are obtained via interpolation between grid points.
}
\label{fig:heatmaps}
\end{figure}


The wound simulations are conducted using overdamped equations of motion (Eq. \ref{suppeq:overdamped}), which are commonly used to model cell dynamics in the viscous extracellular environment \cite{barton2017active, delile2017cell}. We do not model explicit chemical signaling, and instead capture the biomechanical response that results from these chemical signals. Wounds are simulated by first generating a nearly confluent cell monolayer with $\mathcal{A} = 1.2$, similar to $\mathcal{A}$ of embryo and larval wing disc cells. We then remove the central cells from the monolayer, resulting in wounds similar in size to those in laser ablation experiments on epithelial monolayers~\cite{tetley2019tissue}. In these simulations, we focus on the purse-string (PS) mechanism for wound closure. While embryonic wound healing features both PS and protrusive crawling activity \cite{tetley2019tissue,abreu2012drosophila}, our simulations with only PS activity can predict the differences in embryonic and larval wing disc wound closure. We define the PS as a collection of $N_{p}$ vertices along the wound boundary as shown in Fig.~\ref{fig:modelSchematic}d. Each PS vertex at position $\vec{r}_p(t)$ is initially coincident with a wound-adjacent DP vertex at $\vec{r}_d(t)$, such that $\vec{r}_p(0) = \vec{r}_d(0)$. The two vertices are bonded by a spring with stiffness $k_{p}$, length $l = |\vec{r}_{p} - \vec{r}_{d}|$, and yield length $l_y$. For each PS vertex, the interaction energy is
\begin{equation}
    \label{eq:energy_k_PS}
    U_{PS} = \frac{k_p}{2} \left[\min\left(l, \frac{l_y}{2}\right)\right]^2 ~ \Theta(l_y - l),
\end{equation}
where $\Theta(\cdot)$ is the Heaviside step function, and $U_{PS}$ saturates when $l \geq l_y/2$ and vanishes when $l > l_y$. Including $l_y$ ensures that PS vertices only interact with membrane segments near the wound. The PS contracts linearly in time $t$ with constriction rate $\omega$. Adjacent PS vertices are connected by springs with stiffness $k_{ps} = k_l$ and rest length 
\begin{equation}
    \label{eq:PS_rest_length}
    \lambda_0(t) = \lambda_0(0) - \frac{\omega}{N_{p}},
\end{equation}
which causes the PS to constrict over time. $\lambda_0(0)$ is chosen for each PS segment such that there is no initial tension, i.e. $\lambda_0(0) = l_{\alpha i}$. 

To compare the wound closure simulation results with those from experiments, we analyze confocal microscopy images of wound closure in embryo and wing disc epithelia from Ref. \cite{tetley2019tissue}. (See Figs. S3 and S4 in SM \cite{supp}.) We convert simulation units to physical units using estimates of the adhesive force between two cells $f_{adh} (\approx 1~$nN) \cite{krieg2008tensile}, cell area ($a_0 \approx 25~\mu$m$^2$ for embryo ectoderm and $a_0 \approx 16~\mu$m$^2$ for wing disc epithelium), and PS constriction rate ($\omega \approx 0.3~\mu$m/s) \cite{biron2005molecular, stachowiak2014mechanism}. For example, the plastic relaxation time $\tau$ and cell bulk modulus $B$ can be expressed in physical units as 
\begin{equation}
    \label{eq:tau_real_units}
    \tau = \tau^* \sqrt{a_0}/\omega 
\end{equation}
\begin{equation}
    B = k_a^*\frac{f_{adh}}{a_0} \label{eq:B_derivation},
\end{equation}
where $\tau^*$ is the dimensionless plastic relaxation time and $k_a^*$ is the dimensionless area stiffness spring constant. 

By varying $\tau$ and using realistic values of $B$, the wound closure simulations can recapitulate cell shape changes near the wound $\Delta\mathcal{A} = {\cal A} - {\cal A}(0)$ (where ${\cal A}(0) \approx 1.2$) and closure rates $dA/dt$ that mimic those for embryo and wing disc wound closure. In Fig.~\ref{fig:heatmaps}, we show $\mathcal{A}$ of cells in the healed tissue that were adjacent to the wound boundary and $dA/dt$ as a function of $B$ and $\tau$. Increasing cell membrane plasticity (i.e. decreasing $\tau$) significantly increases $dA/dt$ and $\Delta\mathcal{A}$. Elastic-like cells only achieve comparable $\Delta\mathcal{A}$ to embryo cells when they are unrealistically soft ($B \leq 0.04$ kPa), as experimental measurements of cell bulk moduli range from $0.3$ to $2$ kPa \cite{charras2005non,coughlin2006filamin, charras2008life}. In a realistic range of $B$, elastic-like cells feature decreased $dA/dt$ and smaller $\Delta\mathcal{A}$, with final shapes ranging from $\mathcal{A} = 1.35 $ to $1.5$ (Fig. \ref{fig:heatmaps}b). Small $B$ alone does not result in the order of magnitude difference in $dA/dt$ between embryo and wing disc wounds, suggesting that plasticity is essential to achieve $\Delta\mathcal{A}$ and $dA/dt$ found during embryo wound closure.

In elastic-like cells (i.e. $\tau>85$ min in Fig. \ref{fig:heatmaps}), increased $B$ causes decreases in $dA/dt$ and $\Delta\mathcal{A}$. This trend reverses in plastic-like cells (i.e. $\tau < 20$ min in Fig. \ref{fig:heatmaps}). Increasing $B$ dramatically decouples changes in membrane length from changes in area. Therefore, work done by $U_p$ strains membrane lengths significantly more than cell areas, enhancing $\Delta\mathcal{A}$ when the membranes are plastic. We confirm this result by demonstrating that stiffer, plastic cells are more deformable than softer or more elastic cells, in simulations of a cell experiencing an extensile force dipole. (See Fig. S5 in SM \cite{supp}.)

\begin{figure}[t]
    \centering
    \includegraphics[width=\linewidth]{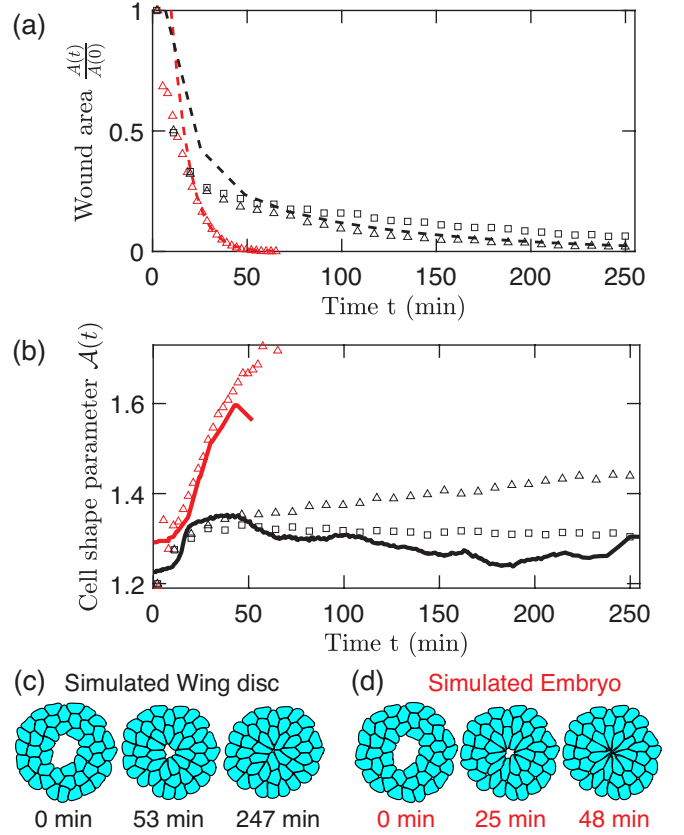}
\caption{(a) Normalized wound area $A(t)/A(0)$ plotted versus time $t$ for simulations of wound closure in embryo (red triangles) and wing disc cells (black triangles) using parameters in Fig. \ref{fig:heatmaps}. (b) $\mathcal{A}(t)$ averaged over cells adjacent to the wound for the simulations in (a). Solid lines indicate the moving average of $\mathcal{A}(t)$ from experiments. Results from simulations using Eq. \ref{eq:tau_s} are included in (a) and (b) to account for shape memory of wing disc cells (black squares). Snapshots of these simulations are shown for (c) embryo and (d) wing disc parameters.
}
\label{fig:validation}
\end{figure}

We validate the DP model for wound closure by comparing the results of simulations in Fig. \ref{fig:heatmaps} to time-series data for the wound area and cell shape parameter in Fig. \ref{fig:experimentalObservations}a,b. The simulation parameters, except $\tau$ and $B$, are identical among different simulations in Fig. \ref{fig:heatmaps}. We use softer and more plastic cells to model the embryo (red square) than the wing disc (black square). This choice reflects the expectation that cell stiffness increases with greater degree of cell differentiation \cite{guimaraes2020stiffness, molnar2021plastic}, and the observation that embryo cell shapes deform faster and more severely than wing disc cell shapes. We find that our results for the time dependence of the wound area are consistent with the experimental data for the embryo and wing disc (Fig. \ref{fig:validation}a). $\mathcal{A}(t)$ in embryos matches the simulation results. In contrast,  $\mathcal{A}(t)$ in wing discs (Fig. \ref{fig:validation}b) requires an additional shape memory term in the membrane remodeling equation, {\it c.f.} Eq.~\ref{eq:plastic_dynamical_eq}, 
\begin{equation}
    \label{eq:tau_s}
    \frac{dl_{0\alpha i}}{dt} = -\frac{k_l}{\eta}\left(l_{0\alpha i}-l_{\alpha i} - \xi(l_{\alpha i} - l_{0\alpha i}(0)\right),
\end{equation}
where $l_{0\alpha i}(0)$ is the segment length before wounding and $\xi$ controls the timescale $\tau_s = \tau/\xi$ for cells to recover their original shape. We use $\xi=0$ for the embryo and $\xi = 0.1$ for the wing disc, which leads to a better description of $\mathcal{A}(t)$. This result suggests that more differentiated cells may have greater shape memory, perhaps to maintain their specialized functions.

Snapshots at several time points in Fig. \ref{fig:validation}c-d show how wound closure trajectories vary with $B$ and $\tau$. Soft and plastic cells form multicellular rosettes, which are common in embryonic wound healing and developmental processes~\cite{wood2002wound, tetley2019tissue, harding2014roles}. Elastic-like cells exhibit less elongation towards the wound and fewer cell contacts near the wound, which is attributed to increased intercalation, akin to the wing disc wound closure process \cite{tetley2019tissue}. The difference in elongation between elastic- and plastic-like cells during wound closure arises due to differences in cell stress relaxation, as plastic cells can relax by elongation whereas elastic cells must relax by changing neighbors.

Deformable particle model simulations show that changes in cell stiffness and membrane plasticity lead to distinct wound closure phenotypes displayed in \textit{Drosophila} embryo ectoderm and larval wing disc epithelium. The simulations take advantage of the deformable particle model's ability to describe highly deformed cell shapes to incorporate membrane plasticity which is not present in previous vertex model simulations of wound closure. By varying the cell bulk modulus $B$ and plastic relaxation timescale $\tau$, we find regimes that correspond to fast closure with significant cell shape changes, and slow closure with minor cell shape changes, which recapitulate the wound healing experiments on embryo and larval wing disc epithelia. We attribute the increased wound closure rates and greater cell shape changes to enhanced cell shape deformability, controlled by $B$ and $\tau$, in embryo cells relative to wing disc cells. These results show that the apparent paradox in Ref. \cite{tetley2019tissue} that embryo ectodermal wounds close more quickly than wing disc epithelial wounds can be resolved by considering a cell model with cell shape plasticity. Our model predicts that the key to explaining the differences in wound closure of \textit{Drosophila} embryo and wing disc wounds is cell shape plasticity.

The correlation between rapid closure rate and distorted cell shapes during epithelial wound healing may suggest that earlier developmental stages prioritize fast healing at the expense of not maintaining the original tissue structure. Future work is necessary to determine whether cell shape plasticity is responsible for distinct patterns of tissue restructuring that occur during developmental processes, such as gastrulation and neurulation \cite{solnica2012gastrulation,vijayraghavan2017mechanics}. In addition, although the current study investigates cell shape changes from membrane surface remodeling, the role of cell volume plasticity is still unclear. Cell shape plasticity is the net result of surface area and volume plasticity, which are influenced by actin cortex remodeling and membrane reservoirs. Future experiments are necessary to investigate the separate contribution of these components to cell shape plasticity during wound closure. 

The authors acknowledge support from NSF Grant No. 2102789, NIH T32 GM 008384, and the High Performance Computing facilities operated by Yale’s Center for Research Computing. We also thank Robert Tetley and Yanlan Mao for sharing their wound healing data.

\appendix
\bibliography{main}

\end{document}


\preprint{APS/123-QED}

\title{Supplemental Material for \\ Mechanical Plasticity of Cell Membranes Enhances Epithelial Wound Closure}

\author{Andrew T. Ton}
\affiliation{
 Department of Physics, Yale University, New Haven, Connecticut 06520, USA}
\affiliation{
 Integrated Graduate Program in Physical and Engineering Biology, Yale University, New Haven, Connecticut 06520, USA}
 
\author{Arthur K. MacKeith}
\affiliation{
 Department of Mechanical Engineering \& Materials Science, Yale University, New Haven, Connecticut 06520, USA}
\affiliation{
 Integrated Graduate Program in Physical and Engineering Biology, Yale University, New Haven, Connecticut 06520, USA}
 
\author{Mark D. Shattuck}
\affiliation{
 Benjamin Levich Institute and Physics Department, The City College of New York, New York, New York 10031, USA}

\author{Corey S. O'Hern}
\affiliation{
 Department of Mechanical Engineering \& Materials Science, Yale University, New Haven, Connecticut 06520, USA}
\affiliation{
 Department of Physics, Yale University, New Haven, Connecticut 06520, USA}
\affiliation{
 Department of Applied Physics, Yale University, New Haven, Connecticut 06520, USA}
\affiliation{
 Integrated Graduate Program in Physical and Engineering Biology, Yale University, New Haven, Connecticut 06520, USA}

\date{\today}

\maketitle

The supplemental material includes seven sections that provide additional details concerning the deformable particle simulations and analyses of the experimental data of wound closure in epithelial monolayers. In Section I, we describe the intercellular forces, initialization of the cell monolayer, generation of the wound, equations of motion, and tracking of the wound boundary over time in the numerical simulations of wound closure. In Section II, we provide a novel method for calculating smooth intercellular forces in the deformable particle model simulations. In Section III, we describe the cell segmentation, cell shape parameter calculations, and error estimation from the analyses of confocal microscopy images of wound healing in epithelial tissues in {\it Drosophila}~\cite{tetley2019tissue}. In Sections IV and V, we relate the cell bulk modulus to the deformable particle area stiffness, and express the bulk modulus and other simulation quantities in physical units. In Section VI, we study the deformation of a cell experiencing an extensile force dipole to give further evidence that a larger bulk modulus leads to enhanced shape changes in plastic cells. In Section VII, we discuss multiple contributing factors to the overall process of cell shape plasticity.

\section{Wound closure simulation protocol}

\begin{figure*}
    \centering
    \includegraphics[width=\linewidth]{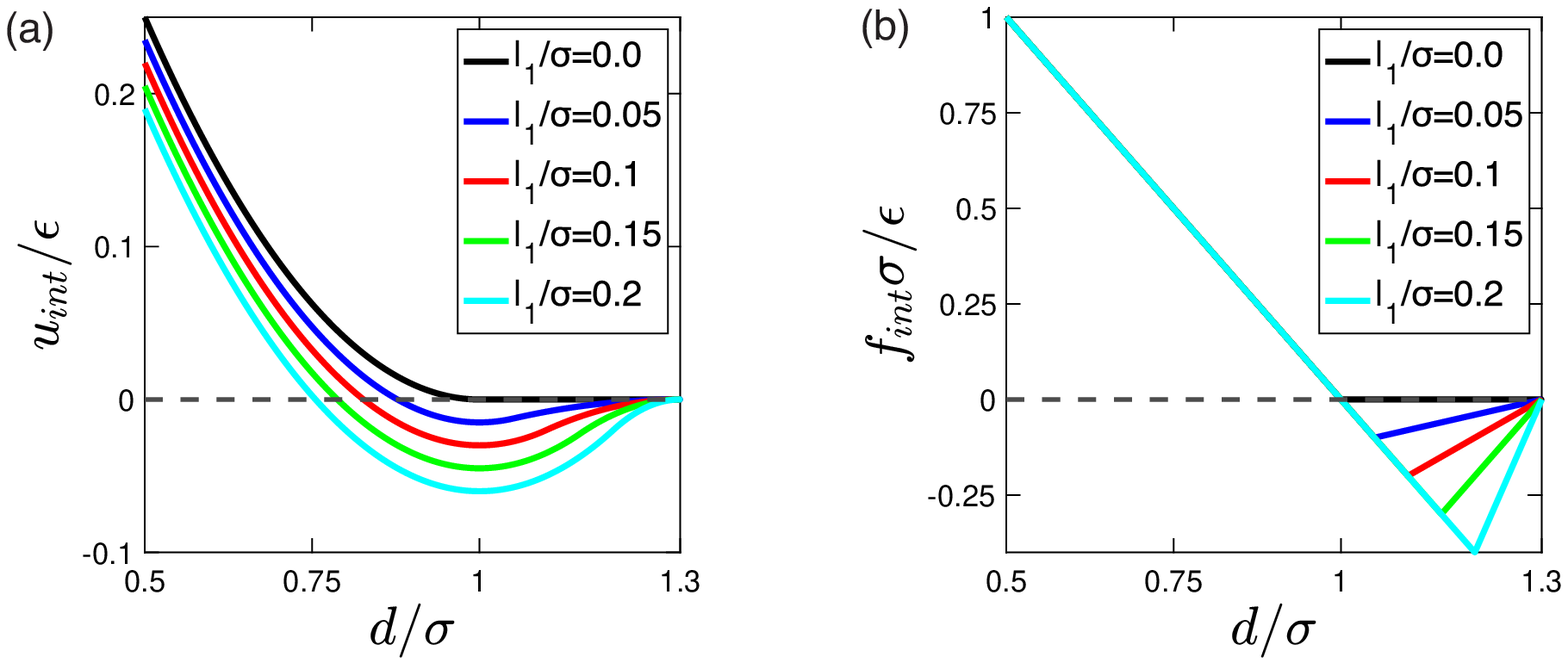}
\caption{(a) Vertex-segment pair potential energy $u_{int}$ and (b) vertex-segment force $f_{int}$ is plotted against the vertex-segment distance $d$ for several values of the attraction strength $l_1/\sigma$. $u_{int}$, $f_{int}$, $d$, and $l_1$ are all nondimensionalized using the half membrane width $\sigma$ and membrane interaction strength $\epsilon$. $l_2$ is fixed at $0.3\sigma$ for all numerical simulations of wound closure.  
}
\label{suppfig:potential}
\end{figure*}

We model cell monolayers as nearly confluent packings of deformable particles. Each deformable particle (cell) $i$ obeys the shape energy in Eq. 1 in the main text and interacts with other cells through vertex-segment forces between vertex $\alpha$ on cell $i$ and segment ${\vec l}_{\beta j}$ on cell $j$ as shown in Fig. 2 in the main text. The intercellular forces can be derived using the following pair potential: 
\begin{align}\label{suppeq:pair_potential}
    U_{int} = \sum_{i>j} \sum_{\alpha > \beta} u_{int}(d_{\alpha i, \beta j}),
\end{align}
where 
\begin{align}
    u_{int}(d) &=  
    \begin{cases}
        \epsilon\left[(1-\frac{d}{\sigma})^2 - \frac{l_1l_2}{\sigma^2}\right], &0 < \frac{d}{\sigma} < 1 + \frac{l_1}{\sigma}\\
        \epsilon \frac{l_1}{l_1-l_2} \left[1+\frac{l_2}{\sigma}-\frac{d}{\sigma}\right]^2, &1 + \frac{l_1}{\sigma} < \frac{d}{\sigma} < 1 + \frac{l_2}{\sigma},
    \end{cases}
\end{align}
$\epsilon$ is the cell membrane interaction strength, $d\equiv d_{\alpha i, \beta j}$ is the vertex-segment distance, $\sigma$ is half the membrane width, $l_2 = 0.3\sigma$ sets the attractive range, $l_1$ controls the membrane attraction strength, and $l_1 < l_2$. The vertex-segment pair potential $u_{int}$ and the force $f_{int} = -\partial{u_{int}}/\partial{d}$ are plotted against $d/\sigma$ in Fig. \ref{suppfig:potential}. See Table \ref{tab:defaultParams} for a list of default parameters and other quantities used in the simulations.

To generate unwounded cell monolayers, we first place cells with random initial positions within a circular boundary at an initial packing fraction $\phi_0=0.78$, and then compress the system in small packing fraction increments $\Delta \phi = 0.005$ until $\phi=0.92$. After each compression step, we use the FIRE algorithm \cite{bitzek2006structural} to minimize the total potential energy of the cells within the circular boundary: 
\begin{equation}
U'_{DP} = \sum_i^N U_{shape,i} + U_{rep} + U_{cb},
\end{equation}
where $U_{rep}$ is equal to $U_{int}$ with $l_1=0$ and
\begin{equation}
U_{cb} = 
\\
\begin{cases}
    \epsilon_{cb}\sum\limits_{i}^{N} \sum\limits_{\alpha}^{N_V} \left(1 - \frac{|r_{\alpha i} - R|}{\sigma}\right)^2, & |r_{\alpha i} - R| < \sigma \\
    
    0, &\text{otherwise}.
\end{cases}
\end{equation}
Here, $r_{\alpha i}$ is the distance between a vertex $\alpha$ on cell $i$ and the center of the circular boundary of radius $R$. We set $\epsilon_{cb} = \epsilon$ as the strength of the interaction energy between the circular boundary and the vertices. 

The resulting packings of deformable particles with ${\cal A}=1.2$ and $\phi=0.92$ serve as initial conditions for the numerical simulations of wound closure. First, we add cell-cell adhesion to the packings by setting $l_1/\sigma = 0.1$ and carry out constant energy dynamics without the circular boundary for a total time $\sim \sqrt{a_0}/\omega$ to allow the cells to explore intercellular gaps. We then minimize the total potential energy $U_{DP}$ (Eq.~4 in the main text) using an overdamped equation of motion, where each vertex $\alpha$ on cell $i$ obeys
 \begin{equation}
    \frac{d\vec{r}_{\alpha i}}{dt} = \frac{\vec{f}_{\alpha i}}{b}.
    \label{suppeq:overdamped}
\end{equation}
Here, $\vec{f}_{\alpha i}$ is the total force on the vertex $i$, and $b$ is the damping coefficient. Overdamped dynamics are frequently employed in active particle models \cite{barton2017active} to balance the input of energy due to activity, which corresponds to the purse-string contraction in this model. The effects of the cells' environment on cell dynamics are commonly modeled using a single damping coefficient \cite{barton2017active, delile2017cell}, as we have done in Equation \ref{suppeq:overdamped}.

We then introduce a wound by removing 5 central cells in the packing, define the purse-string (PS) on the wound boundary, and integrate Eq. \ref{suppeq:overdamped} using a modified velocity-Verlet algorithm with timestep $dt^* = 0.05/\sqrt{k_a^*}$. We record cell properties, such as the cell shape parameter $\mathcal{A}$, and wound area $A$ throughout the course of the simulation. The simulated wound boundary is tracked over time by monitoring vertex-vertex contacts and determining the largest cluster of vertices near the center of the tissue using the Newman-Ziff union-find algorithm \cite{newman2001fast}. We terminate the wound closure simulation when the wound area satisfies $A < 10^{-2} a_0$. In both the numerical simulations and image analyses, a cell is considered wound-adjacent if it is within $\sqrt{a_0/\pi}$ of the wound boundary.

\section{Smooth sliding intercellular forces}

\begin{figure*}
    \centering
    \includegraphics[width=\linewidth]{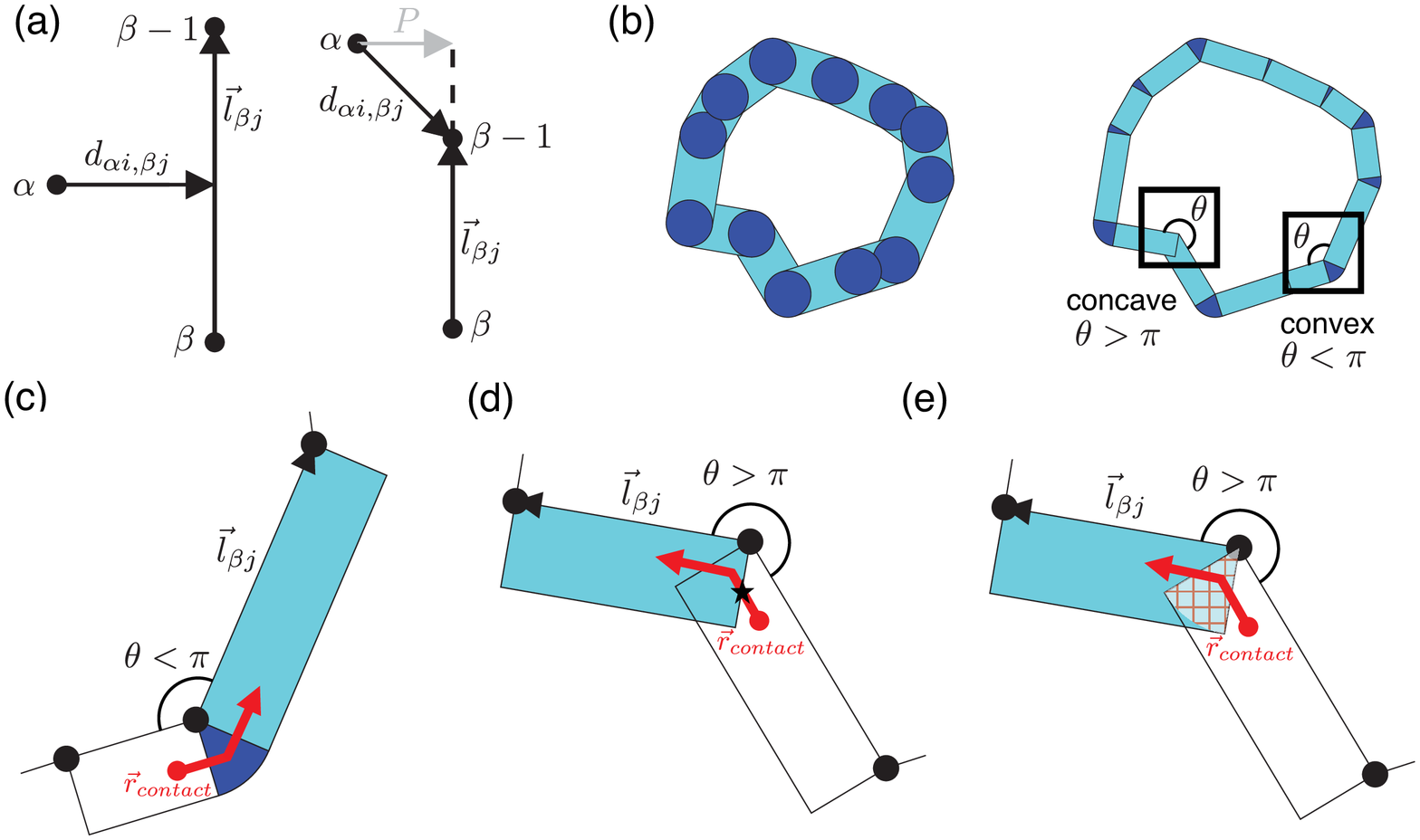}
\caption{(a) The vertex-segment distance $d_{\alpha i, \beta j}$ is the shortest distance from vertex $\alpha$ on cell $i$ to line segment ${\vec l}_{\beta j}$ on cell $j$. If the projection $P$ of ${\vec d}_{\alpha i, \beta j}$ to the segment ${\vec l}_{\beta j}$ (gray arrow) falls outside of $l_{\beta j}$, then $d_{\alpha i, \beta j}$ is the distance between vertex $\alpha$ on cell $i$ and vertex $\beta$ or $(\beta-1)$ on cell $j$. (b) Each deformable particle (left) is modeled as a collection of vertices (dark blue circles) connected by line segments (cyan region). In the right image, we also show the exterior-facing half of the cell membrane of a ``smooth" deformable particle, which consists of circulo-line segments (cyan) and vertex sectors (blue). Within a smooth deformable particle, the section around each vertex is considered concave when the angle the vertex makes with its two neighboring vertices has an interior angle $\theta > \pi$, and convex otherwise. (c) For locally convex geometries, when a vertex that overlaps the membrane surface at $\vec{r}_{contact}$ moves along the indicated trajectory (red line), the  vertex-segment distance relative to ${\vec l}_{\beta j}$ is continuous. (d) For locally concave geometries, the vertex-segment distance (relative to ${\vec l}_{\beta j}$) for a vertex overlapping with the membrane along the trajectory $\vec{r}_{contact}$ (red line) undergoes a discontinuous jump when crossing the point indicated by the black star. (e) The wedge-shaped patch (red grid) indicates the region over which the force in Eq. \ref{suppeq:vertexPatchForce} acts to remove the discontinuity in force that occurs in $\partial u_{int}/\partial d_{\alpha i, \beta j}$ in the case of locally concave membrane geometries. The wedge also removes a similar discontinuity relative to $l_{(\beta - 1)j}$.
}
\label{suppfig:smooth}
\end{figure*}

Intercellular forces are bumpy in the deformable particle model when the intercellular potential is a function of the vertex-vertex distances. We develop a novel method (within the deformable particle model) to compute \textit{smooth} (frictionless) sliding intercellular forces. To do this, we assume that the intercellular pair potential $U_{int}$ is only a function of the closest distance $d_{\alpha i, \beta j}$ between vertex $\alpha$ on cell $i$ and line segment ${\vec l}_{\beta j}$ on cell $j$ (Eq.~3 in the main text), which ensures that there is no component of the intercellular force tangential to the membrane at $\vec{l}_{\beta j}$ as shown in Fig. \ref{suppfig:smooth}a.  This force law mimics smooth surfaces that consist of connected rectangles (cyan) and wedges (blue) in Fig. \ref{suppfig:smooth}b. Because $d_{\alpha i, \beta j}$ is a function of $\vec{r}_{\alpha i}$, $\vec{r}_{\beta j}$, and $\vec{r}_{(\beta-1) j}$, forces computed using the pair potential $u_{int}(d_{\alpha i,\beta j})$ will affect the dynamics of all three vertices $\alpha$, $\beta$, and $\beta-1$.

For interaction potentials that depend on the closest distance between a point and a line segment, one must determine whether there are discontinuities in the force that can occur when the closest distance changes discontinuously even though a vertex or line segment moves by an infinitesimal amount. We consider two cases, concave and convex sections of the cell membrane, classified by the interior angle $\theta$ defined by three successive vertices. In the convex case, $\theta < \pi$ (Fig. \ref{suppfig:smooth}c), a vertex that overlaps with the membrane surface at $\vec{r}_{contact}$ can slide along the surface with a continuous vertex-segment distance $d_{\alpha i,\beta j}$ to the line segment ${\vec l}_{\beta j}$. In the concave case, $\theta > \pi$ (Fig. \ref{suppfig:smooth}d), a vertex on a similar trajectory will experience a discontinuity in the vertex-segment distance, since the concave surface (Fig. \ref{suppfig:smooth}b) lacks a wedge where there is an overlap of the two rectangles. One method to remove this discontinuity is to add a wedge-shaped patch, shown as a red grid in Fig. \ref{suppfig:smooth}e. Within the wedge-shaped region, a vertex $\alpha$ overlapping with the membrane surface $\l_{\beta j}$ at $\vec{r}_{contact}$ incurs an additional force 
\begin{equation}
    f_{patch} = +\frac{\partial u_{int}}{\partial d_{\alpha i, \beta j}},
    \label{suppeq:vertexPatchForce}
\end{equation}
such that in the patch region, $f_{patch}$ provides an equal and opposite force to offset the discontinuity in $f_{int}(d_{\alpha i, \beta j})$ (Fig. \ref{suppfig:potential}b) that occurs when the vertex $\alpha$ enters the patch region. 

\section{Image analysis}
\begin{figure*}
    \centering
    \includegraphics[width=\linewidth]{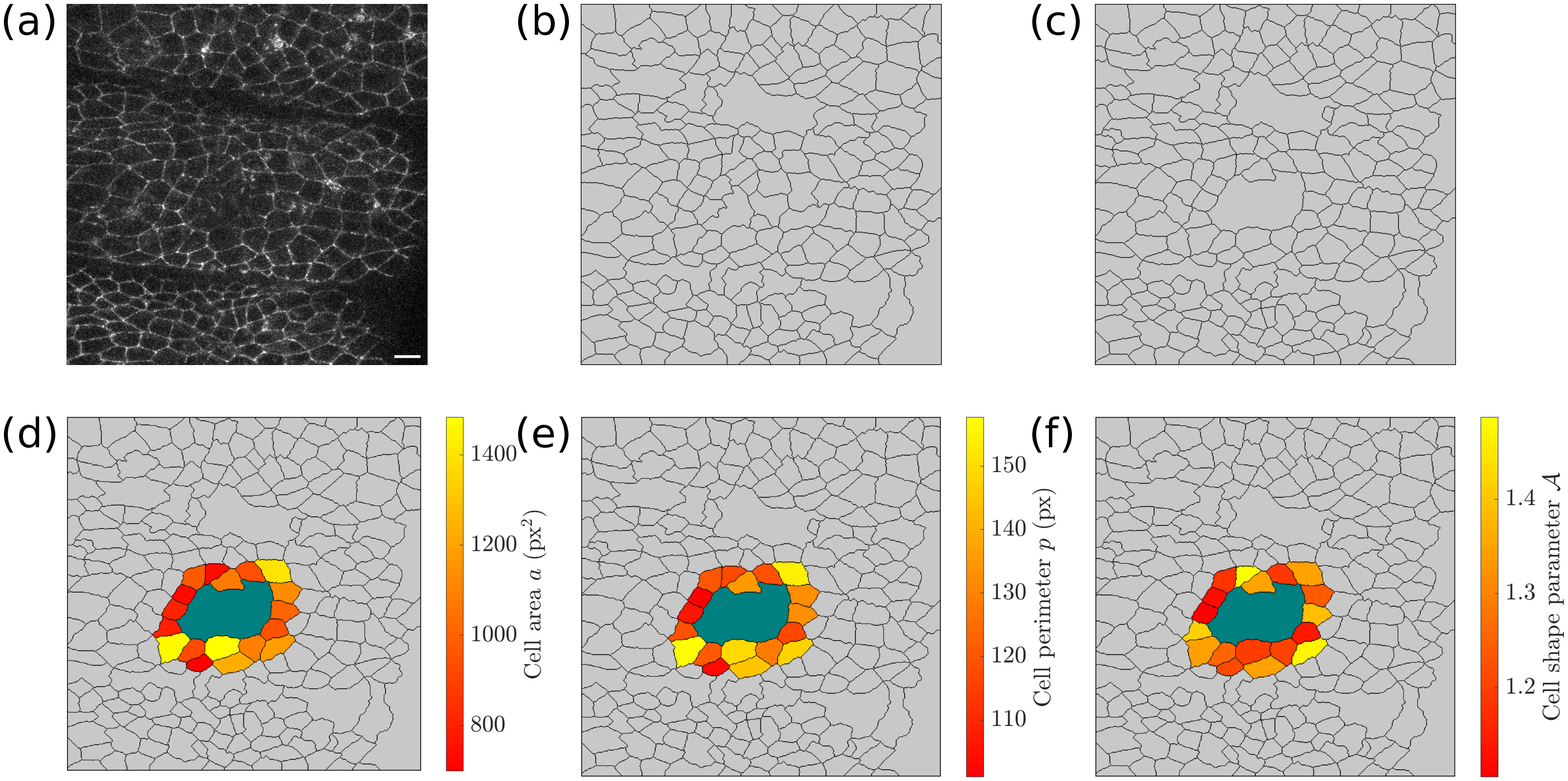}
\caption{(a) The first step in the image analysis pipeline is to transform raw confocal microscopy images into maximum intensity projections along the z-axis, shown here for a \textit{Drosophila} embryo ectoderm immediately after wounding from Ref.~\cite{tetley2019tissue}. The scale bar has width 10 $\mu$m. (b) We next perform automated cell boundary segmentation using Tissue Analyzer \cite{Aigouy2016}, an ImageJ plugin for segmentation of single-layered epithelia. (c) We make manual corrections to cell boundaries near the wound in each frame. From the segmented cell boundaries, we calculate (d) the cell area $a$ and the wound area, (e) the cell perimeter $p$, and (f) the shape parameter $\mathcal{A}$ of cells adjacent to the wound. In this example, the wound area is $9048$ px$^2$ or $157$ $\mu$m$^2$.}
\label{suppfig:imagAnalysisPipeline}
\end{figure*}

To measure the cell shape parameters and wound area over time of the wounded embryo and wing disc epithelia, we analyze segmented images of the wound closure process from Ref. \cite{tetley2019tissue}. We use the segmented cell boundaries in the $5$ wing disc wounds found in Ref. \cite{tetley2019tissue}. For the $2$ embryo wounds in Ref.~\cite{tetley2019tissue}, we perform our own segmentation procedure by first taking a maximum intensity projection (Fig. \ref{suppfig:imagAnalysisPipeline}a) along the $z$-axis of confocal microscope $z$-stack time-series images. Next, we use \textit{Tissue Analyzer} \cite{Aigouy2016}, an ImageJ plugin for segmentation of single-layered epithelia, to obtain a first pass segmentation of the cells and wound (Fig. \ref{suppfig:imagAnalysisPipeline}b). Then we make manual corrections to account for under- and over-segmented regions near the wound, which yields a binary image of the cell boundaries (Fig. \ref{suppfig:imagAnalysisPipeline}c). We employ \textit{regionprops} in MATLAB R2022a, Update 5 (Fig. \ref{suppfig:imagAnalysisPipeline}d-e) to calculate the area and perimeter of each unique segmented region, which we use to report the wound areas and cell shape parameters $\mathcal{A}$ for each time point. 

\begin{figure*}
    \centering
    \includegraphics[width=\linewidth]{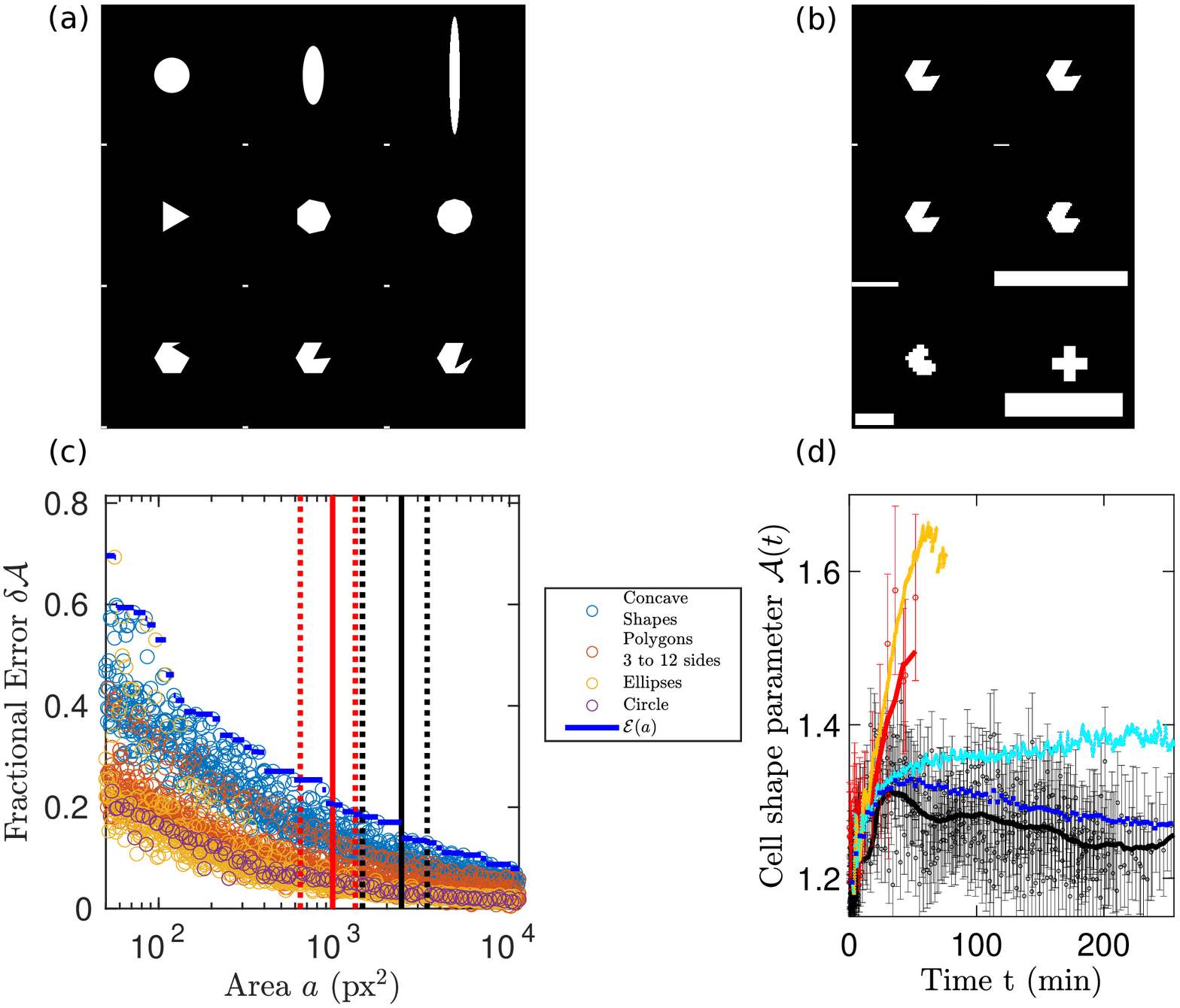}
\caption{(a) Synthetic data is shown at high resolution: From left to right, then top to bottom: ellipse with $e=0$ (circle), ellipse with $e=0.933$, ellipse with $e=0.996$, triangle, heptagon, dodecagon, concave shape $1$, $4$, and $8$. The scale bars to the lower left of each shape are 100 px wide. (b) A shape with an interior angle $\theta > \pi$ is shown at several resolutions. The scale bars to the lower left of the top four images are 100 px wide, and the scale bars in the bottom two are 10 px wide. The scale bar sizes in the images vary due to significant changes in the image resolutions. (c) The fractional error in $\mathcal{A}$ measured in synthetic data is plotted versus the number of pixels contained inside the shape over an experimentally relevant range. Estimate of the error in $\mathcal{A}$ as a function of the area of the shape in the synthetic images (blue line). The red and black solid lines represent the mean areas in the experimental data sets of embryo and wing disc cells, respectively, and the spacings between the dotted lines represent the standard deviations. (d) We plot $\mathcal{A}(t)$ for the embryo (red) and wing disc (black) experiments using the variance-weighted mean (open circles) over cells adjacent to the wound, where the variance is given by $\mathcal{E}^2(a)$. We include error bars using the standard error of the weighted mean. The solid lines are moving averages with the same window size as in Fig. 1b. We overlay the simulation results for the embryo (yellow triangles), the wing disc (cyan triangles), and the wing disc using cells with shape memory (blue squares), all of which are the same data as in Fig. 4 in the main text.
}
\label{suppfig:imagAnalysisErrors}
\end{figure*}

To generate error estimates for the cell shape parameter measurements, we carry out a similar process on synthetic data. We use $26$ test shapes: ellipses with eccentricities $0.996$, $0.987$, $0.968$, $0.933$, $0.872$, $0.768$, $0.586$, and $0$ (circle), polygons with $3$ to $12$ sides, and $8$ different $7$-sided concave shapes. A subset of the test shapes is shown in Fig. \ref{suppfig:imagAnalysisErrors}a. We generate images of these shapes with a range of resolutions (Fig. \ref{suppfig:imagAnalysisErrors}b) and compare the measured $\mathcal{A}$ to the true value ${\mathcal A}_t$ for each shape (Fig. \ref{suppfig:imagAnalysisErrors}c). We define the fractional error of $\mathcal{A}$ as
\begin{equation}
    \delta \mathcal{A} = \frac{\mathcal{A}_{t} - \mathcal{A}}{\mathcal{A}_{t}}.
    \label{suppeq:fractional_error}
\end{equation}

For a cell with area $a$, we report an estimate of $\delta\mathcal{A}$ using the bounding fractional error $\mathcal{E}(a)$ as shown in Fig. \ref{suppfig:imagAnalysisErrors}c. $\mathcal{E}(a)$ is calculated by taking the maximum $\delta\mathcal{A}$ over all synthetic test shapes at each area, and storing the running maximum as a function of decreasing area. For each measurement of the cell shape parameter $\mathcal{A}$ for a cell with area $a$ (px$^2$) in the embryo and wing disc wounding experiments, we associate $\mathcal{E}(a)$ with the measurement error in ${\mathcal A}$. 

We plot $\mathcal{A}(t)$ taking the variance-weighted mean over cells adjacent to the wound boundary (Fig. \ref{suppfig:imagAnalysisPipeline}d), with error bars given by the standard error of this weighted mean. Onto the variance-weighted mean and error bars, we overlay the simulation results as in Fig. 4b in the main text to show that $\mathcal{A}(t)$, from simulations of the embryo and the wing disc using cells with shape memory, falls within the margin of error for the $\mathcal{A}(t)$ from the experimental measurements of the embryo and wing disc wounds. 

\section{Derivation of Bulk Modulus}
The bulk modulus $B$ of a single cell is related to the cell area stiffness $k_a$ through Eq. 9 in the main text. To derive this relation, we start with the definition 
\begin{align*}
    B&=-V\frac{dP}{dV},
\end{align*}
where $P$ is the pressure and $V$ is the cell volume. Rearranging, we obtain
\begin{align*}
    dP &= -\frac{B}{V} dV,
\end{align*}
and $P = B \log(V/V_0)$, where $V = V_0$ at zero pressure. The energy under isothermal compression is given by  
\begin{align*}
    U_{c} = \int PdV' &= BV(\log(\frac{V}{V_0} - 1) + BV_0.
\end{align*}
To second order in $V/V_0 - 1$, we obtain
\begin{align}
    U_c \approx \frac{1}{2}BV_0\left(\frac{V}{V_0} - 1\right)^2.
    \label{suppeq:quadratic_U}
\end{align}
Comparing Eq. \ref{suppeq:quadratic_U} to the shape-energy function (Eq. 1 in the main text), the energy due to compression in two dimensions is given by 
\begin{align*}
    U_{DP, c} = \frac{1}{2} k_a a_0^2 \left(\frac{a}{a_0} - 1\right)^2.
\end{align*}
Assuming that the energy scale of volume changes is equal to that of area changes, and that $V_0 = a_0^{3/2}$, 
\begin{align*}
    B = k_a\sqrt{a_0}.
\end{align*}

\section{Converting simulation units to physical units}
Simulation units can be converted into physical units using three physical quantities that set the mass, length, and time scales of the simulation. We are able to determine these scales using a choice of force, velocity, and area (See Table \ref{tab:defaultParams}.). Atomic force microscopy can determine single cell forces \cite{kashef2015quantitative}, which allows us to estimate the unsticking force $f_{adh} = 1~$nN based on cohesion between Zebrafish embryo ectodermal cells \cite{krieg2008tensile}. We are unaware of any measurements of $f_{adh}$ on \textit{Drosophila} embryo ectoderm, and we assume that the measurements of $f_{adh}$ on Zebrafish embryo ectoderm give a reasonable estimate. We choose $\omega = 0.3~\mu $m/sec based on a typical actin ring constriction rate~\cite{biron2005molecular, stachowiak2014mechanism}. We find that typical \textit{Drosophila} cell areas are $a_0 \sim 16~\mu$m$^2$ for the late-stage larval wing disc epithelium and $\sim 25~\mu$m$^2$ for the late-stage embryo ectoderm.

We define the unsticking force in our simulations by $f_{adh}=l_1\epsilon N_v/3\sigma^2$ with units of $\epsilon/\sigma$. A factor of $N_v/6$ comes from the assumption that two cells are adhered to each other through $1/6$ of their membranes on average, given that a cell has approximately $6$ neighbors. The maximum vertex-vertex adhesion force in Eq. \ref{suppeq:pair_potential} is $2l_1\epsilon/\sigma^2$, again with units of $\epsilon/\sigma$.  

\section{Stiffness enhances shape change in plastic cells}
\begin{figure*}
    \centering
    \includegraphics[width=\linewidth]{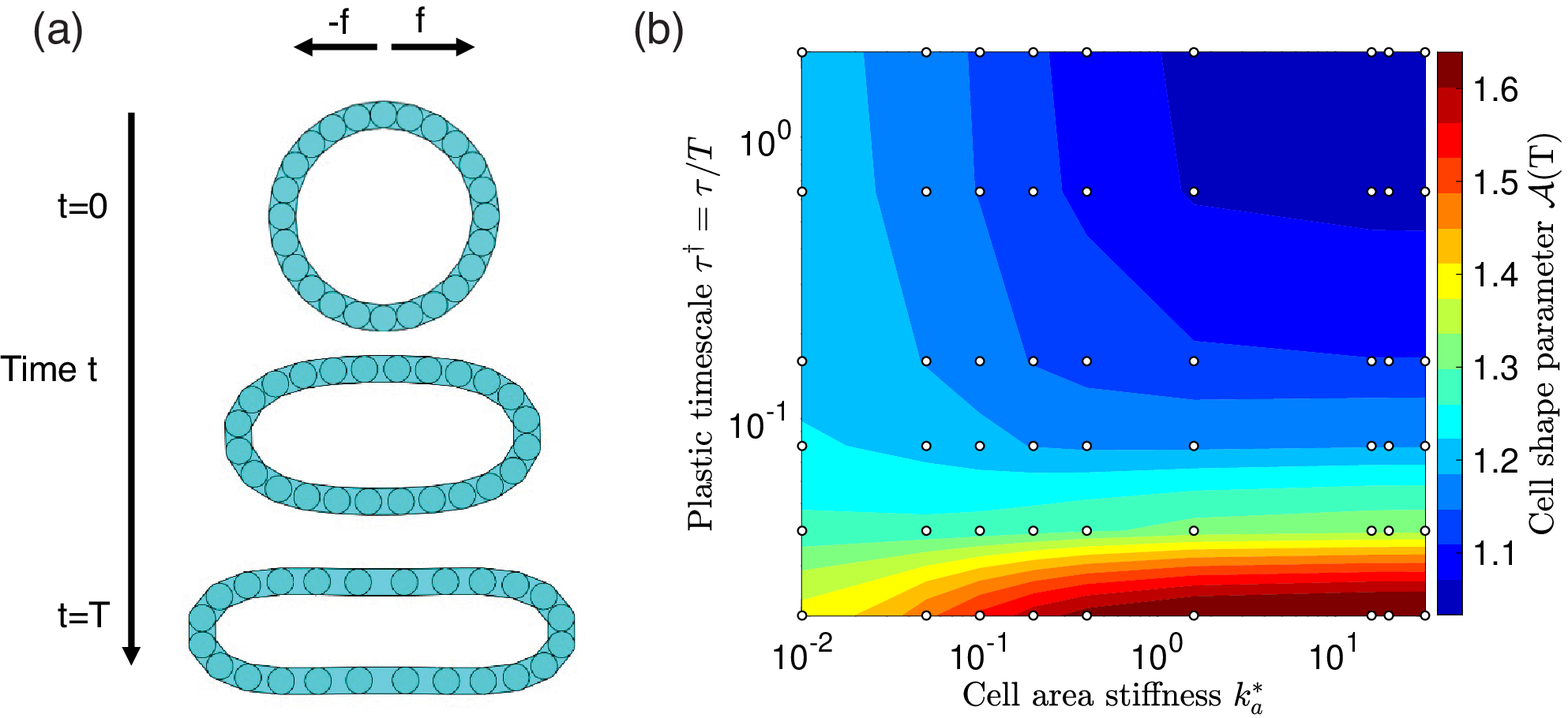}
\caption{(a) A deformable particle is shown with an extensile force dipole with magnitude $f$ that generates elongation over time $t$. (b) Final cell shape $\mathcal{A}(T)$ of the extensile deformable particle as a function of the cell area stiffness $k_a^*$ and the normalized plastic relaxation timescale $\tau^\dag = \tau / T$, where $T$ is the duration of the stretching simulation. 
}
\label{suppfig:forceDipole}
\end{figure*}

To understand how greater stiffness can lead to enhanced shape change in plastic cells, we conduct simulations varying the plastic relaxation timescale $\tau$ and the cell area stiffness $k_a$ of a single cell experiencing an extensile force dipole (Fig. \ref{suppfig:forceDipole}a), i.e. the cell experiences two equal and opposite forces that generate net zero force. For elastic-like cells with large $\tau$, we find that increasing $k_a$ leads to a reduction in the final $\mathcal{A}$ (Fig. \ref{suppfig:forceDipole}b), which matches the expectation that stiffer cells are less deformable. However, the trend reverses for plastic-like cells with small $\tau$, as increasing $k_a$ leads to an increase in the final $\mathcal{A}$. We find that stiff, plastic cells are able to lengthen their membranes without changing their area. In contrast, soft plastic cells increase their area when lengthening their membranes, which results in more modest shape changes. These results show that plastic cells become more deformable as they become stiffer, and corroborates a similar trend in Fig. 3 in the main text. 

\section{Alternative mechanisms influencing cell shape plasticity}
Cell shape plasticity, the property of cells allowing them to retain their new shapes after deformation, is the result of several mechanisms, which include actin cortex remodeling, caveolae acting as membrane reservoirs, and vesicle trafficking processes like exocytosis and endocytosis. These processes are involved in mechanoprotection, and allow the cell to modulate the cell membrane surface area in response to stress. In our model, a natural way to incorporate changes in cell membrane surface area due to membrane reservoirs and vesicle trafficking is to add plasticity in the membrane rest length. We describe cell shape plasticity as membrane plasticity using Equation \ref{eq:plastic_dynamical_eq}. A different model for cell shape plasticity could add plasticity in the equilibrium bending angle of each membrane segment, which would account for how actin cortex remodeling contributes to cell shape plasticity independently of membrane surface area relaxation processes. Since curvature is dependent on both membrane segment lengths and the angles between the segments, we note that membrane rest length plasticity also includes relaxation of membrane curvature. 

\begin{table*}[t]
  \centering
  \begin{tabularx}{\textwidth}{|X|X|X|}
    \hline
    \textbf{Simulation quantities} & \textbf{Symbol} & \textbf{Value} \\
    \hline
    \textbf{Embryo cell rest area} &  $\bf{a_0}$ & $25$ $\mu$m$^2$ \\ 
    \hline
    \textbf{Wing disc cell rest area} & $\bf{a_0}$ & $16$ $\mu$m$^2$ \\
    \hline
    \textbf{Unsticking force} & $\bf{f_{adh}}$ & $1$ nN \cite{krieg2008tensile}\\ 
    \hline
    \textbf{PS constriction rate} & $\bm{\omega}$ & $0.3$ $\mu$m/sec \cite{biron2005molecular, stachowiak2014mechanism}\\ 
    \hline
    Numerical integration timestep & $dt^*$ & $0.1 l_0^*(0) \sqrt{a_0^*/k_a^*}$\\ 
    \hline
    DP vertex damping coefficient & $b^*$ & 1 (damped) \\ 
    & & 0 (constant energy)\\
    \hline
    Number of vertices per DP & $N_v^*$ & $30$ \\ 
    \hline
    Cell area stiffness & $k_a^*$ & $0.25$, $0.5$, $1.0$,$\ldots$,$256$ \\
    \hline
    Membrane length spring constant & $k_l^*$ & $1$ \\ 
    \hline
    Membrane bending rigidity & $k_b^*$ & $0.01$ \\ 
    \hline
    Cell rest area & $a_0^*$ & $1$ \\
    \hline
    Initial membrane segment rest length & $l_0^*(0)$ & $\sqrt{\mathcal{A}4\pi a_0^*}/N_v$ \\ 
    \hline
    Half membrane width & $\sigma^*$ & $l_0^*(0)/2$ \\ 
    \hline
    Membrane interaction energy & $\epsilon^*$ & $1$ \\ 
    \hline
    Maximum vertex-vertex adhesion force & $2l_1^*$ & $0.2$ \\ 
    \hline
    Unsticking force & $f_{adh}^* = l_1^* N_v/3$ & $1$ \\ 
    \hline
    Plastic relaxation timescale & $\tau^* = \eta^* / k_l^*$ & $2.4$, $4.8$, $9.6$,$\ldots$,$39322$ \\ 
    \hline
    PS-DP spring constant & $k_p^*$ & $4$ \\ 
    \hline
    PS-DP spring yield length & $l_y^*$ & 4$\sigma^*$ \\ 
    \hline
    PS constriction rate & $\omega^*$ & $1$ \\ 
    \hline
    Wound closure rate & $\omega \sigma dA^*/dt^*$ & --- ($\mu$m/s)\\ 
    \hline
    Cell shape parameter & $\mathcal{A}$ & ---\\ 
    \hline
  \end{tabularx}
  \caption{Default parameters and other quantities used in the deformable particle simulations of wound closure. An asterisk denotes non-dimensional simulation units. Bolded quantities have physical units and are based on experimental measurements. If only one parameter value is listed, then the parameter is not varied in our simulation studies. }
  \label{tab:defaultParams}
\end{table*}

\clearpage
\bibliography{main}